# Spin-Glass-like Transition and Hall Resistivity of $Y_{2-x}Bi_xIr_2O_7$

Natsuki Aito, Minoru Soda, Yoshiaki Kobayashi and Masatoshi Sato

*Department of Physics, Division of Material Science, Nagoya University,*
*Furo-cho, Chikusa-ku, Nagoya 464-8602*

Various physical properties of the pyrochlore oxide $Y_{2-x}Bi_xIr_2O_7$ have been studied. The magnetizations $M$ measured under the conditions of the zero-field-cooling(ZFC) and the field-cooling(FC) have different values below the temperature $T=T_G$. The anomalous $T$-dependence of the electrical resistivities ρ and the thermoelectric powers $S$ observed at around $T_G$ indicates that the behavior of the magnetization is due to the transition to the state with the spin freezing. In this spin-frozen state, the Hall resistivities $\rho_H$ measured with the ZFC and FC conditions are found to have different values, too, in the low temperature phase ($T<T_G$). Possible mechanisms which induce such the hysteretic behavior are discussed.

corresponding author : M. Sato (e-mail : msato@b-lab.phys.nagoya-u.ac.jp)

## 1. Introduction

Pyroclore oxides $A_2B_2O_7$ have the structure composed of individual networks of the corner-sharing $A_4$ and $B_4$ tetrahedra.[1)] This characteristic structure often introduces the magnetic frustration, and anomalous physical behaviors have been reported in various pyrochlore systems.[2-4)] $Y_{2-x}Bi_xRu_2O_7$[3)] and $Sm_{2-x}Ca_xRu_2O_7$[4)] exhibit the Mott insulator to metal transition with increasing $x$. In $Y_{2-x}Bi_xRu_2O_7$, the metal-insulator (M-I) transition is induced by the change of the transfer energy or the Ru $t_{2g}$ band width, as indicated by the $x$-dependence of the low temperature specific heat coefficient $\gamma$. The M-I transition in $Sm_{2-x}Ca_xRu_2O_7$ is, on the other hand, caused by the change of the carrier number. In both systems, the spin-glass-like transition and the specific heat jump have been observed at $T_G$ in the region of small $x$ and the significant increase of the Hall resistivity $\rho_H$ has also been found with decreasing temperature $T$ above $T_G$. Neutron Rietveld analyses of $Y_2Ru_2O_7$ and $Nd_2Ru_2O_7$ revealed that the ordering of the Ru-magnetic moments is almost long-range in the spin-glass-like phase.[5)]

Pyrochlore iridates are different from the pyrochlore ruthenates in the point that $Ru^{+4}$ ions in the Mott insulating state of the ruthenates have the spin $s$=1, while $Ir^{4+}$ of the latter is expected to have $s$=1/2. Then, the magnetic anisotropy does not exist and effects of the spin fluctuations are expected to be more significant in the iridates. One of the iridate systems, $Y_2Ir_2O_7$ is a Mott insulator and with substitution of Ca for Y, the M-I transition takes place as in the ruthenium pyrochlore systems.[6,7)]

In the present paper, we report results of studies on $Y_{2-x}Bi_xIr_2O_7$, which exhibits the insulator to metal transition with increasing $x$[8)] by the same mechanism as that for $Y_{2-x}Bi_xRu_2O_7$. Detailed data of the transport and magnetic properties, indicate that the system undergoes the transition to the spin-frozen state at $T=T_G$ with decreasing $T$ in the region of small $x$. The anomaly of the specific heat $C$ at $T_G$ observed for $x$=0 is found to be much smaller than that of $Y_2Ru_2O_7$.

The Hall resistivities $\rho_H$ below the temperature~$T_G$ measured by the four probe method exhibit hysteretic behavior, that is, the values taken under the conditions of the zero-field-cooling(ZFC) and the field-cooling(FC) have been found to be different. Discussion on the mechanism of this hysteretic behavior is presented, including the one proposed by Tatara and Kawamura[9)] that the uniform spin chirality induced by the external magnetic field in the spin-glass phase contributes to the Hall resistivity.

## 2. Experiments

Mixtures of $Y_2O_3$, $Bi_2O_3$ and Ir with proper molar ratios were ground, pressed into pellets and heated at 1000 °C for 24h. Then, they were cooled in the furnace. The processes of the grinding and sintering were repeated once more for the samples with nonzero $x$. For the sample with $x$=0, the processes were repeated many times. The X-ray diffraction studies confirmed that all the samples are single phase. The lattice parameters $a$ calculated from the diffraction data are plotted in Fig.1. The $a$-value increases smoothly with $x$, which guarantees that the substitution of Bi for Y is properly carried out. The electrical resistivities ρ were measured by a conventional four-probe method with an ac-resistance bridge. The magnetizations $M$ or the magnetic susceptibilities χ were obtained by a SQUID magnetometer, where both conditions of the zero-field-cooling(ZFC) and the field-cooling(FC) were used. The specific heat($C$) measurements have been carried out by the thermal relaxation method in the region from 2.5 K to 300 K. Measurements of $C$ in the region 0.9 K<$T$<3 K have also been carried out for $Y_2Ir_2O_7$ by the adiabatic heat pulse method and the low temperature data are published elsewhere.[8)] The thermoelectric powers $S$ were measured by a dc method. In the measurements of the Hall resistivities $\rho_H$, the four probe method was adopted, where the samples were rotated to invert the direction of the applied magnetic field($H$=5 T) to remove the background voltage. The Hall voltages were measured under the conditions of ZFC and FC with increasing and decreasing $T$ stepwise, respectively. Details of the measurements can be found in ref.10.

## 3. Results and Discussion

Figure 2 shows the ρ-$T$ curves of $Y_{2-x}Bi_xIr_2O_7$ for various $x$ values. $Y_{2-x}Bi_xIr_2O_7$ exhibits the insulator to metal transition with increasing $x$. Above 150 K, the metallic $T$-dependence is observed for $x$ larger than ~0.2. For the samples with 0.2≤ $x$<0.5, ρ has a minimum at around the temperature slightly larger than 100 K. It is due to the transition to the low temperature phase, which is also indicated by the observed $T$-dependence of the magnetizations $M$ or the magnetic susceptibilities χ(see Fig. 3). The $x$-region in which the low temperature phase exists, is wider than that observed in $Y_{2-x}Ca_xIr_2O_7$.[7)]

$M$-$T$ curves taken with $H$=1 T for several samples of $Y_{2-x}Bi_xIr_2O_7$ are shown in Fig. 3. The significant difference between the data, $M_{ZFC}$ and $M_{FC}$, taken under the conditions of ZFC and FC, respectively, has been found below the temperatures $T_G$. The temperatures seem to correspond to those of the resistivity minimum. The inset shows the H-dependence of the magnetization measured at 5 K for x=0.3, where the existence of the very small spontaneous magnetization $M_s$ is clearly observed. This $M_s$ has been found to appear at $T_G$ with decreasing T. Although the existence of $M_s$ has been known in $Y_{2-x}Bi_xRu_2O_7$ [3] and other pyrochlore ruthenates,[11] its origin is still unclear.

Figure 4 shows the specific heat data of the samples with x=0.0 and 0.2 against T in the region from 2.5 K to 350 K. At $T_G$, the specific heat C of $Y_2Ir_2O_7$ exhibits the slight anomaly, which is much smaller than that of $Y_2Ru_2O_7$,[3,5] while for x=0.2, the anomaly can hardly be detected. The result is in clear contrast to that of ruthenates, which is one of significant distinctions between ruthenium (s=1) and iridium (s=1/2) pyrochlore systems.

In Fig. 5, the T-dependence of the thermoelectric powers S is shown for several samples of $Y_{2-x}Bi_xIr_2O_7$. For x=0.0 and 0.1, clear anomalies can be observed at the temperatures corresponding to $T_G$. Because the thermoelectric power S is, generally speaking, rather insensitive to the boundaries, the data present clear evidence for the phase change of the present system at $T_G$. It is also supported by the anomalous T-dependence of the resistivity ρ, though ρ is rather sensitive to the grain boundaries. Then, the anomalies observed at $T_G$ can be considered to originate from an intrinsic phase change, possibly to the spin frozen state. One might claim that because just the existence of the spontaneous magnetization $M_s$ produces the observed difference between the magnetizations $M_{ZFC}$ and $M_{FC}$, the difference does not necessarily indicate that the transition is to the spin frozen state. However, by following considerations, we think that the transition at $T_G$ is to the spin-frozen state: The decrease of $M_{ZFC}$ observed below $T_G$ with decreasing T for small values of x, can be considered to be the characteristic of the spin freezing transition and cannot be understood by the existence of nonzero $M_s$. Moreover, the T-dependence is rather similar to that observed in the spin-freezing transition of pyrochlore ruthenates.[3]

It should be added here that the present transition to the spin frozen state is accompanied by the transport anomalies. At x=0, the specific heat anomaly has been observed, which is not expected in the ideal spin-glass transition. (The transition in $Y_2Ru_2O_7$ exhibits much larger specific heat anomaly.) In this sense, the transition is not the ideal spin-glass transition. We call it the spin-glass-like transition.

Magnetization data used in the analyses of the Hall resistivity $\rho_H$ were collected in both directions of H(H=5 T) and –H, where the temperature T was changed stepwise with the field in the normal (not reversed) direction. (Note that in the measurements of $\rho_H$, the samples were rotated by 180° with the field direction being fixed.) Figure 6 shows the T-dependence of $M_{ZFC}$ and $M_{FC}$ measured with H=5 T. The values of $M_{FC}$, $M\uparrow$ and $M\downarrow$, measured at fixed $T$ with $\boldsymbol{H}$ along the normal and reversed directions, respectively, exhibit significant difference below $T_G$, while these values of $M_{ZFC}$ are almost equal to each other in the whole $T$-range. If we write $M\uparrow$ and $M\downarrow$ as $M_{av}+\delta M$ and $M_{av}-\delta M$, respectively, using the average value $M_{av} \equiv (M\uparrow+M\downarrow)/2$, $\delta M$ is considered to be the component which is fixed to the sample and does not move even when the field $\boldsymbol{H}$ is reversed. It is induced along the direction of $\boldsymbol{H}$ during the field cooling. The component $M_{av}$ is the one which changes or moves when $\boldsymbol{H}$ is changed at fixed $T$. In the case of ZFC, $M\uparrow \cong M\downarrow$ or $M_{ZFC} \cong M_{av}$ holds and $\delta M \cong 0$. It is noted here that the most (or dominant) part of $\delta M$ is $M_s$. It should be also noted that the contribution related to the magnetization $\delta M$ to the Hall resistivity $\rho_H$ cannot be observed in the present experiment, because the Hall voltage is determined by taking the difference between the transverse voltages which appear before and after the sample rotation, where the contribution of the magnetization fixed to the sample is cancelled.

In Fig. 7, results of the $\rho_H$-measurements carried out with $H$=5 T are shown in both cases of FC and ZFC, in the forms of $\rho_H$ -$T$, $\rho_H/M_{av}$-$T$ and $\rho_H/\rho^2 M_{av}$-$T$ in the left, center and right columns respectively, for the samples with $x$=0.28, 0.3 and 0.4. In the left panel, we find that the Hall resistivities $\rho_H$ measured under the conditions of ZFC and FC are different below the temperature $T_G$, at which $M\uparrow$ and $M\downarrow$ of $M_{FC}$ begin to separate with decreasing $T$(see Fig. 6). Above $T_G$, $\rho_H$ increases with decreasing $T$.

For ruthenium pyrochlore oxides, similar behavior of $\rho_H$ was reported except for the significant difference between $\rho_H$-values in the ZFC and FC cases (In their studies, only the ZFC condition was used.), where the $T$-dependence of $\rho_H$ was attributed to that of the ordinary Hall coefficient $R_0$[4]: The significant $T$-dependence of $\rho_H/H$ above $T_G$ was considered to be due to the strong antiferromagnetic fluctuations, as in the normal states of high-$T_c$ superconductors.[12,13] The decrease of $\rho_H/H$ below $T_G$ with decreasing $T$ was considered to be due to the suppression of spin fluctuations in the spin-glass phase. The mechanism may be able to explain the present observations, too. However, there remains a problem if the antiferromagnetic fluctuations depend on the cooling conditions or not.

Then, we try to write the Hall resistivity as $\rho_H = R_0 H + 4\pi R_s M$, where $R_s$ is the anomalous Hall coefficient. As stated above, only the component $M_{av}$ contributes to $\rho_H$ in the present measurements. Because the relation $M_{av} \propto \chi H$ holds with $T$-insensitive $\chi$ (see Figs. 3 and 6), the $T$-dependence of $\rho_H/M_{av}$ ($=R_0/\chi+4\pi R_s$) is mainly determined by that of $R_0$ or $R_s$. Then, the question is if there is a possibility, apart from the model which considers effects of the antiferromagnetic fluctuations, that the term $4\pi R_s$ predominantly contributes to the observed hysteretic behavior of $\rho_H$. To study such a case, we plot the $\rho_H/M_{av}$ values in the center column of Fig. 7. If the anomalous $T$-dependence of $\rho_H/M_{av}$ which becomes significant at around $T_G$ with decreasing $T$ is due to the second term of the equation, it is from the anomalous behavior of $R_s$. The conventional theories of the anomalous Hall effect predict that $\rho_H$ has the terms proportional to $\rho M_{av}$ and $\rho^2 M_{av}$.[14,15] Then, we show $\rho_H/\rho^2 M_{av}$ as an extreme case in Fig. 7, where not only the temperatures at which the hysteretic behavior of $\rho_H$ begins with decreasing $T$ but also those of the maximum of $\rho_H/\rho^2 M_{av}$ roughly coincide with $T_G$. Even if $\rho_H \propto \rho$ as another theory proposes,[15] the result does not change much. After all, conventional theories of the anomalous Hall resistivities cannot explain the presently observed anomalous change of the $T$-dependence by the anomalous $T$-dependence of $\rho$ and $M_{av}$. They cannot explain the hysteretic behavior, either.

It is interesting to adopt the idea proposed by Tatara and Kawamura[9] as one of possible mechanisms to explain the observed behavior of $\rho_H$. According to them, the uniform spin chirality $\kappa_0$ induced by the magnetic field or the magnetization produces an excess contribution to the Hall conductivity in the spin-glass phase. (The spin chirality $\kappa$ is defined for three spins $\boldsymbol{s}_1$, $\boldsymbol{s}_2$ and $\boldsymbol{s}_3$ as $\kappa \equiv \boldsymbol{s}_1 \cdot (\boldsymbol{s}_2 \times \boldsymbol{s}_3)$.) Due to this contribution, which appears in the magnetic field as the system goes into the spin-frozen state with decreasing $T$, the anomalous $T$-dependence of $\rho_H/M_{av}$ shows up. The hysteretic behavior of $\rho_H$ may also be explained as the difference between the chiralities induced in the ZFC and FC cases.

In the previous papers,[10,16-19] we reported unusual behavior of the anomalous Hall resistivity found for several systems such as $Cu_{1-x}Zn_xCr_2Se_4$, $CuCr_2S_4$ and $Nd_2Mo_2O_7$ with non-trivial spin ordering and $SrFe_{1-x}Co_xO_{3-\delta}$ and $Fe_{0.7}Al_{0.3}$ which exhibit the ferromagnetic to (reentrant) spin-glass transition. We think that in some of these systems, at least, the unusual behavior of the Hall resistivity is induced by the mechanism related with the spin chirality. Then, even in the present $Y_{2-x}Bi_xIr_2O_7$, it is very plausible that this new mechanism is relevant to the observed anomalous behavior of the Hall resistivity.

## 4. Conclusions

The results of the transport, thermal and magnetic properties in $Y_{2-x}Bi_xIr_2O_7$ have been reported. The hysteretic behaviors are observed in the magnetization and the Hall resistivity below $T_G$ for the samples with small $x$. At around $T_G$, $\rho$, $S$ and $C$ exhibit the anomalous $T$-dependence, which indicates that at $T_G$ the bulk phase transition takes place to the spin-glass-like phase. The anomaly of the specific heat observed in the present system at $T_G$ is much smaller than in the ruthenium pyrochlores. This is a meaningful difference between iridates with $s$=1/2 and ruthenates with $s$=1. The anomalous behaviors of the Hall resistivity are reported and arguments are presented on the possible mechanisms of the anomalous behavior related to the spin fluctuations and the spin chirality.

## 5. Acknowledgement

We wish to thank Prof. Kawamura for helpful comments on the anomalous Hall resistivity from the viewpoint of the chirality mechanism. This work has been supported by Grants-in-Aid for Scientific Research from the Japan Society for the Promotion of Science (JSPS) and from the Ministry of Education, Culture, Sports, Science and Technology.

References

1) M. A. Subramanian, G. Aravamudan and G. V. Subba Rao: Prog. Solid State Chem. **15** (1983) 55.
2) A. P. Ramirez, A. Hayashi, R. J. Cava, R. Siddharthan and B. S. Shastry: Nature **399** (1999) 333.
3) S. Yoshii and M. Sato: J. Phys. Soc. Jpn. **68** (1999) 3034.
4) S. Yoshii, K. Murata and M. Sato: J. Phys. Chem. Solids **62** (2001) 129.
5) M. Ito, Y. Yasui, M. Kanada, H. Harashina, S. Yoshii, K. Murata M. Sato, H. Okumura and K. Kakurai: J. Phys. Soc. Jpn. **69** (2000) 888.
6) N.Taira, M. Wakeshima and Y. Hinatsu: J. Phys. Cond. Mat. **13** (2001) 5527.
7) H. Fukazawa and Y. Maeno: J. Phys. Soc. Jpn. **71** (2002) 2578.
8) M. Soda, N. Aito, Y. Kurahashi, Y. Kobayashi and M. Sato: Physica B to be published.
9) G. Tatara and H. Kawamura: J. Phys. Soc. Jpn. **71** (2001) 2613.
10) T. Kageyama, N. Aito, S. Iikubo and M.Sato: J. Phys. Soc. Jpn. Submitted.
11) N. Taira, M. Wakeshima, and Y. Hinatsu: J. Solid State Chem. **144** (1999) 216.
12) H. Kontani, K. Kanki and K. Ueda: Phys. Rev. B **59** (1999) 14723.
13) T. Nishikawa, J. Takeda and M. Sato: J. Phys. Soc. Jpn. **63** (1994) 1441.
14) L. Berger: Phys. Rev. B**2** (1970) 4559.
15) V. K. Dugaev, A. Crepieux and P. Bruno, Phys. Rev. B**64** (2001) 104411.
16) S. Yoshii, S. Iikubo, T. Kageyama, K. Oda, Y. Kondo, K. Murata and M. Sato: J. Phys. Soc. Jpn. **69** (2000) 3777.
17) T. Kageyama, S. Iikubo, S. Yoshii, Y. Kondo, M. Sato: J. Phys. Soc. Jpn. **70** (2001) 3006.
18) S. Iikubo, Y. Yasui, K. Oda, Y. Ohno, Y. Kobayashi, M. Sato and K.Kakurai: J. Phys. Soc. Jpn. **71**(2002) 2792.
19) T. Ido, Y. Yasui and M. Sato: J. Phys. Soc. Jpn. **72** (2002) No. 2.

Figure captions

Fig. 1 Lattice parameter of $Y_{2-x}Bi_xIr_2O_7$ is plotted against *x*.

Fig. 2 Temperature dependence of the electrical resistivities ρ of $Y_{2-x}Bi_xIr_2O_7$.

Fig. 3 Temperature dependence of the magnetizationof $Y_{2-x}Bi_xIr_2O_7$ measured with *H*=1 T. For the samples with the hysteretic behavior, the larger and smaller values indicate $M_{FC}$ and $M_{ZFC}$, respectively. Inset shows the field dependence of *M* for *x*=0.3 measured at *T*=5K.

Fig. 4 Temperature dependence of the specific heat for *x*=0.0(solid circles) and 0.2(open circles) in the region from 2.5 K to 300 K.

Fig. 5 Temperature dependence of the thermoelectric powers of $Y_{2-x}Bi_xIr_2O_7$.

Fig. 6 Temperature dependence of the magnetization of $Y_{1.7}Bi_{0.3}Ir_2O_7$ measured with *H*=5 T by inverting the direction of the magnetic field. Solid and open circles represent the data obtained under the ZFC and FC conditions, respectively, and *M*↑ and *M*↓ correspond to the measuring field ***H*** and –***H***, respectively(For ZFC, *M*↑ ≅ *M*↓ holds.).

Fig. 7 Temperature dependence of $\rho_H$, $\rho_H/M_{av}$ and $\rho_H/\rho^2 M_{av}$ deduced from the data taken with *H*=5 T is shown, where $M_{av}$ is the average of the magnetizations measured before and after inverting the direction of the magnetic field. Solid and open circles represent the data obtained under the ZFC and FC conditions, respectively.

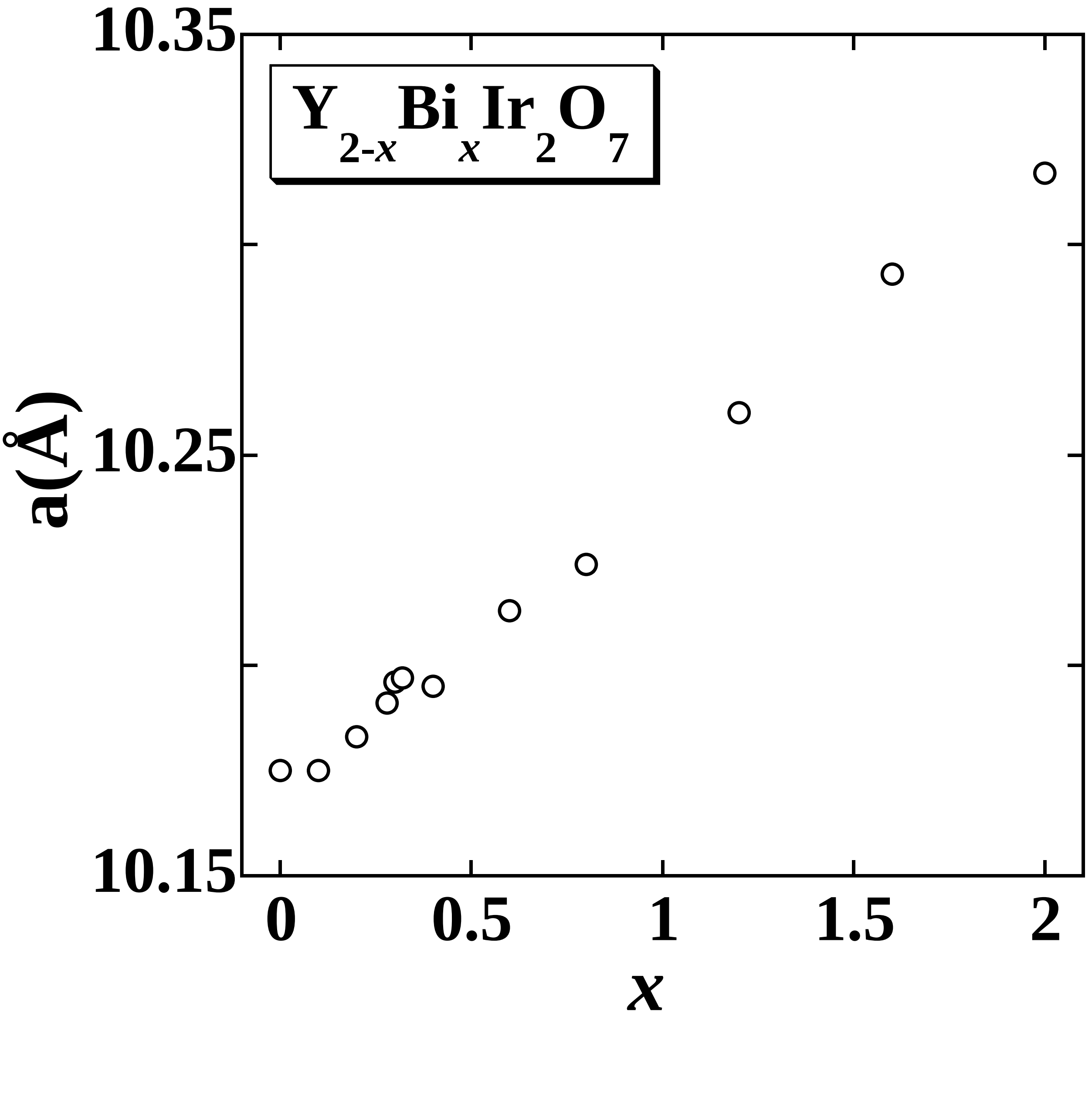

Fig.1

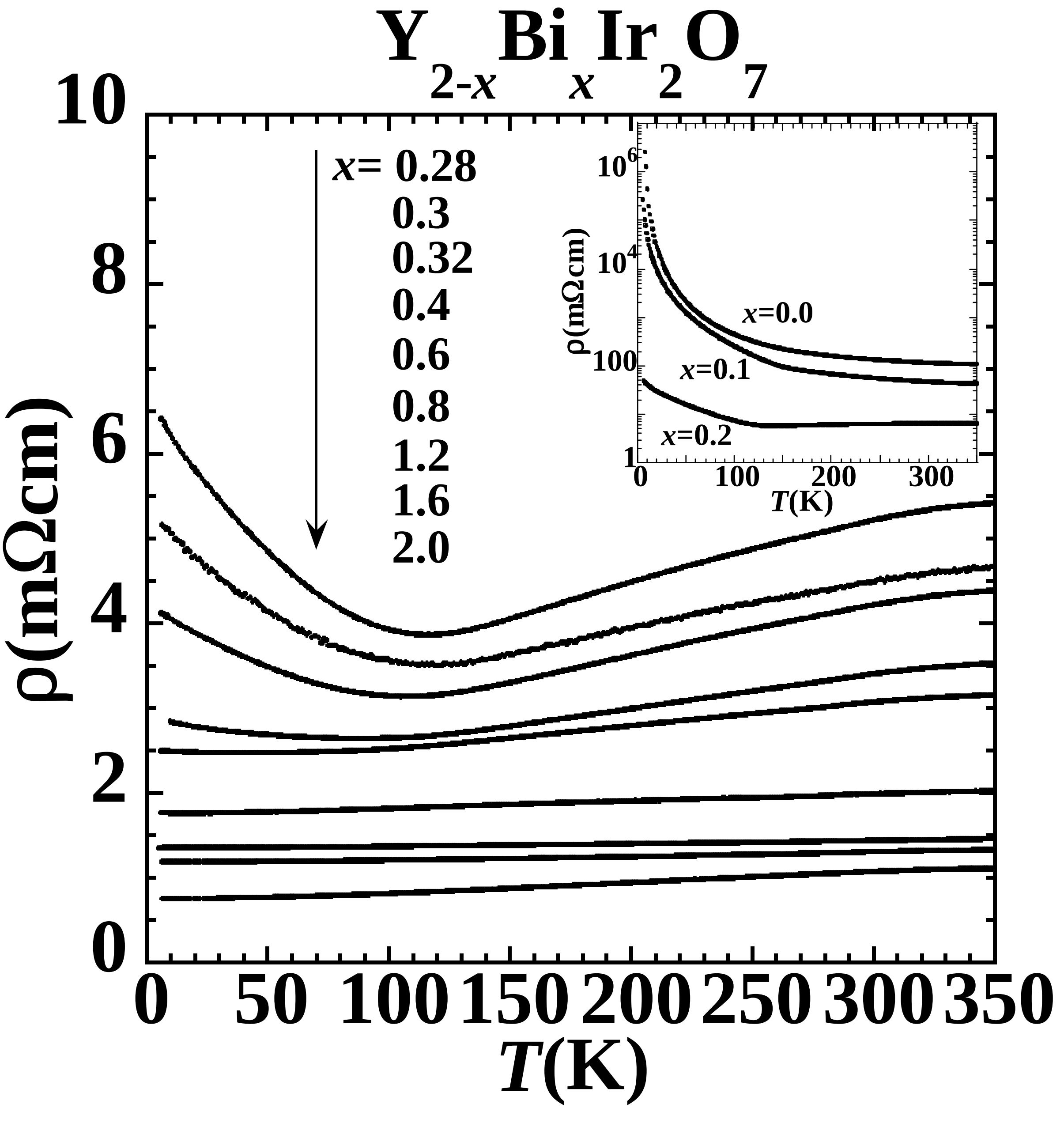

**Fig.2**

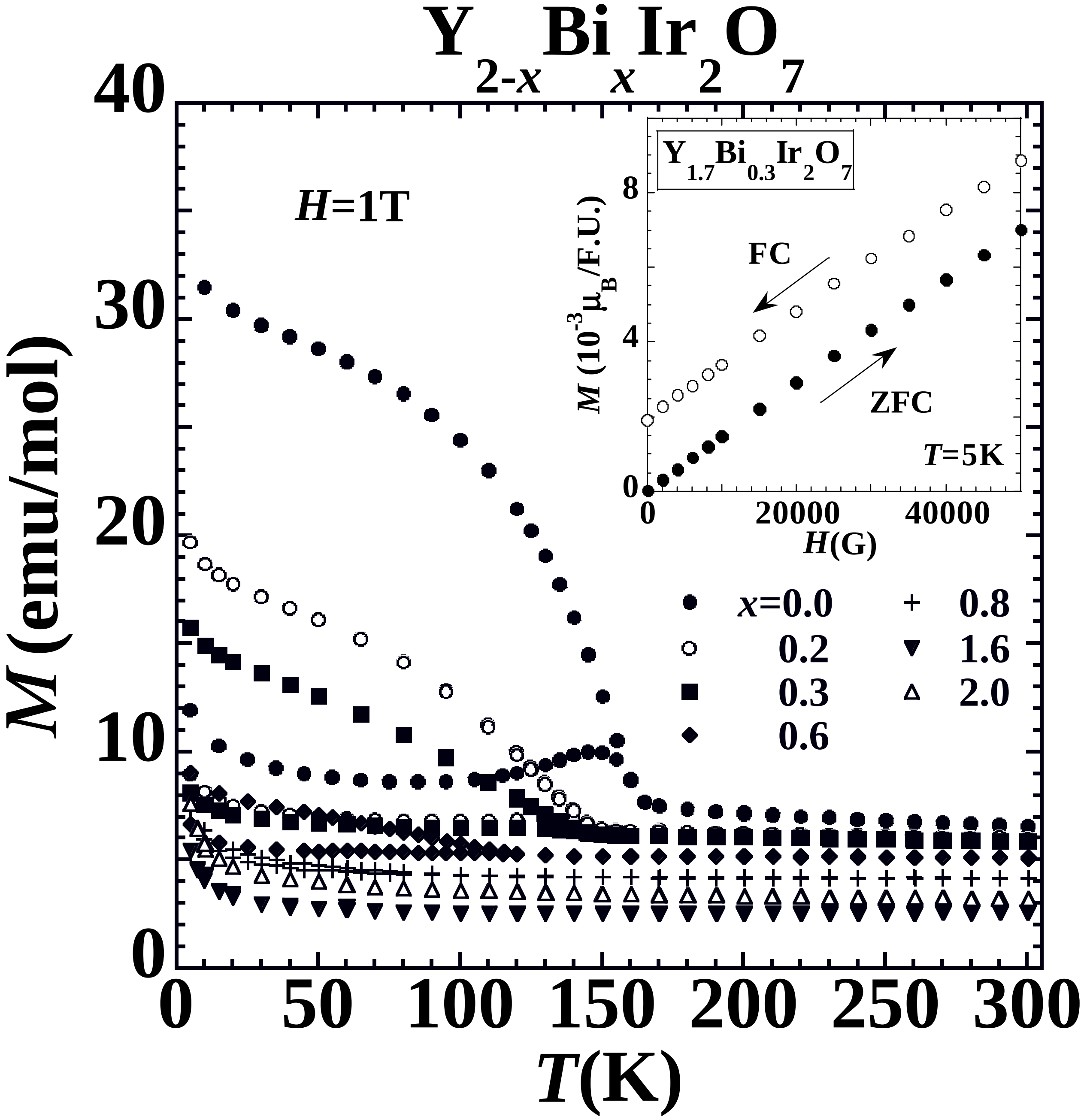

Fig.3

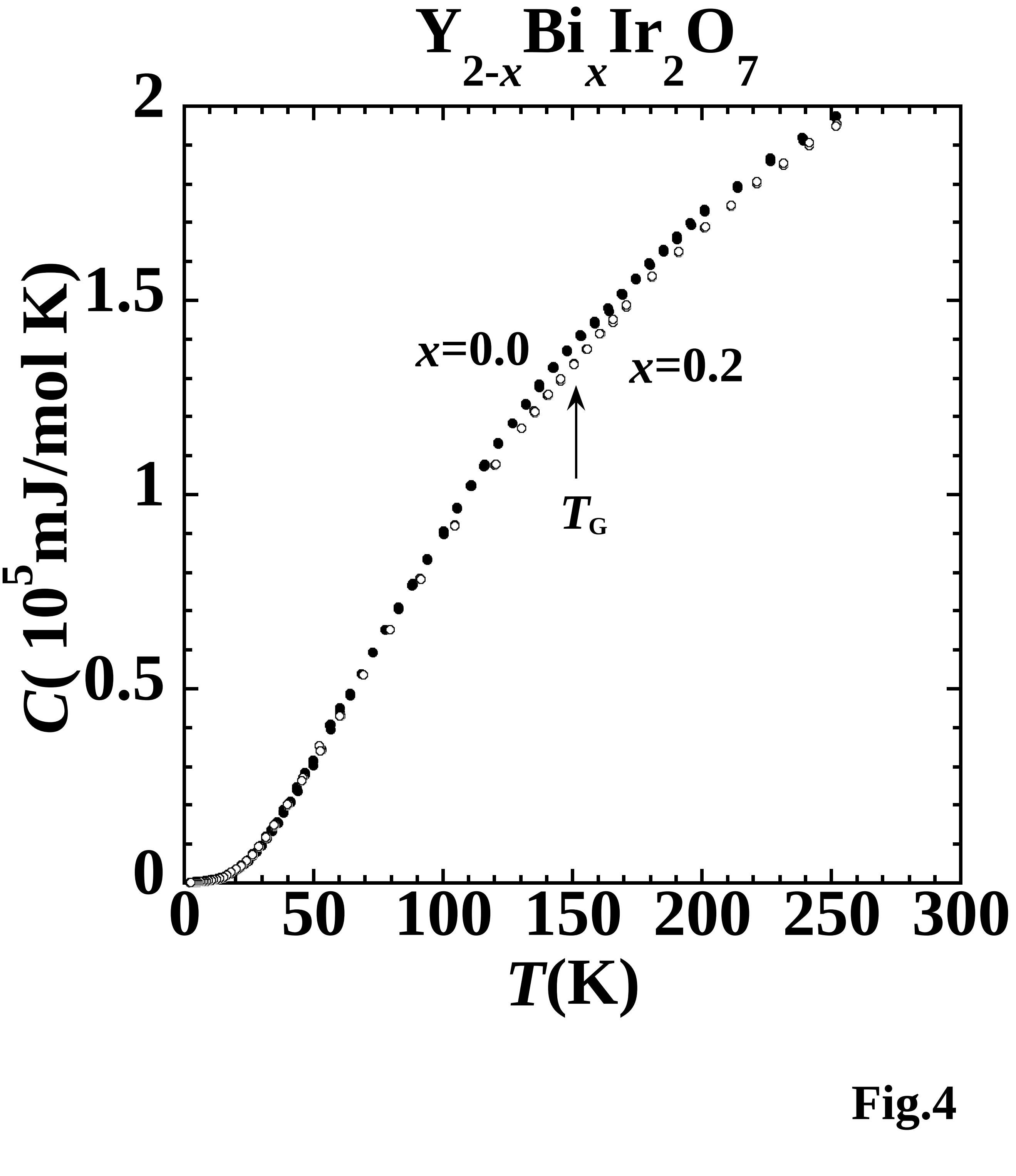

Fig.4

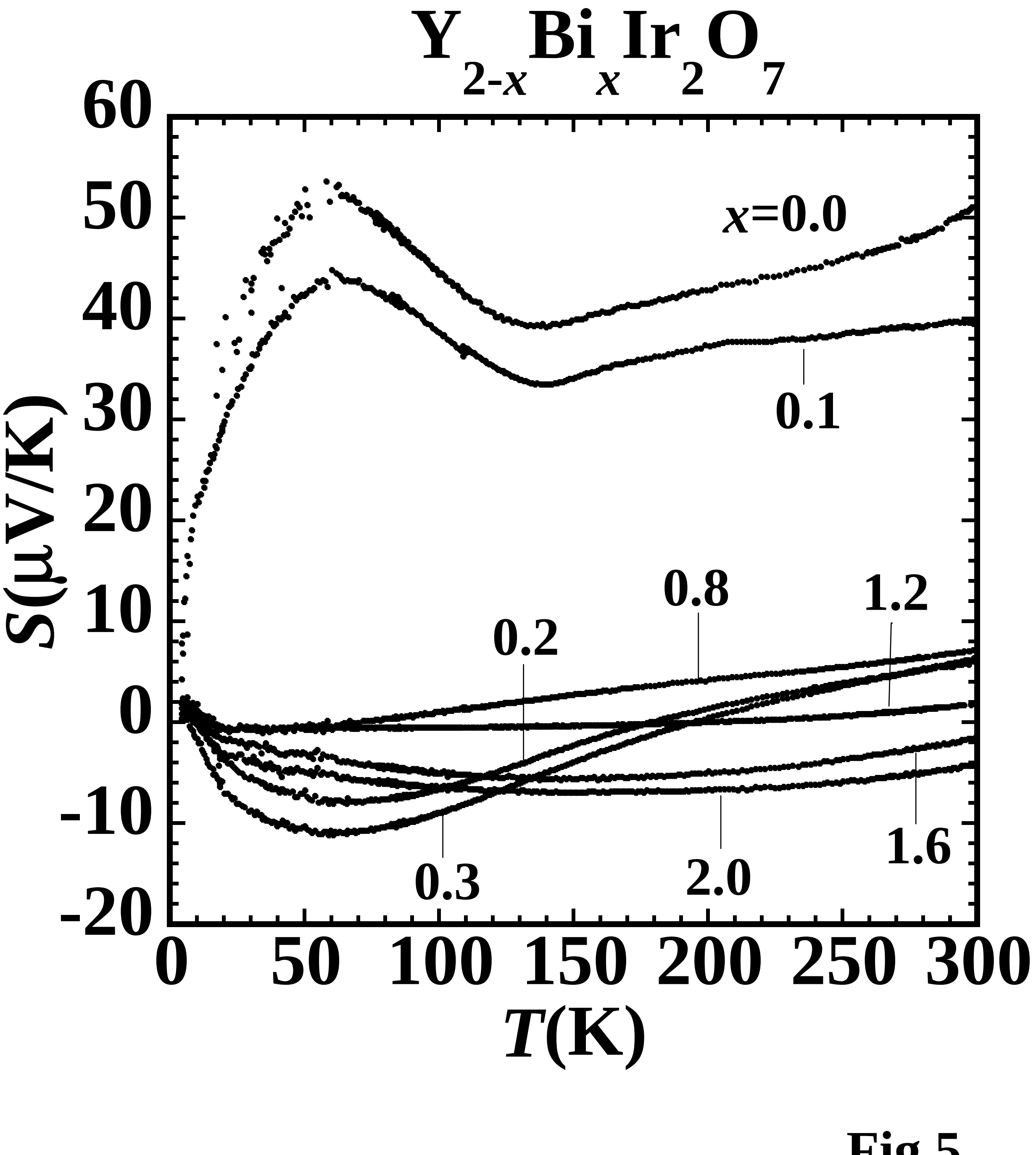

Fig.5

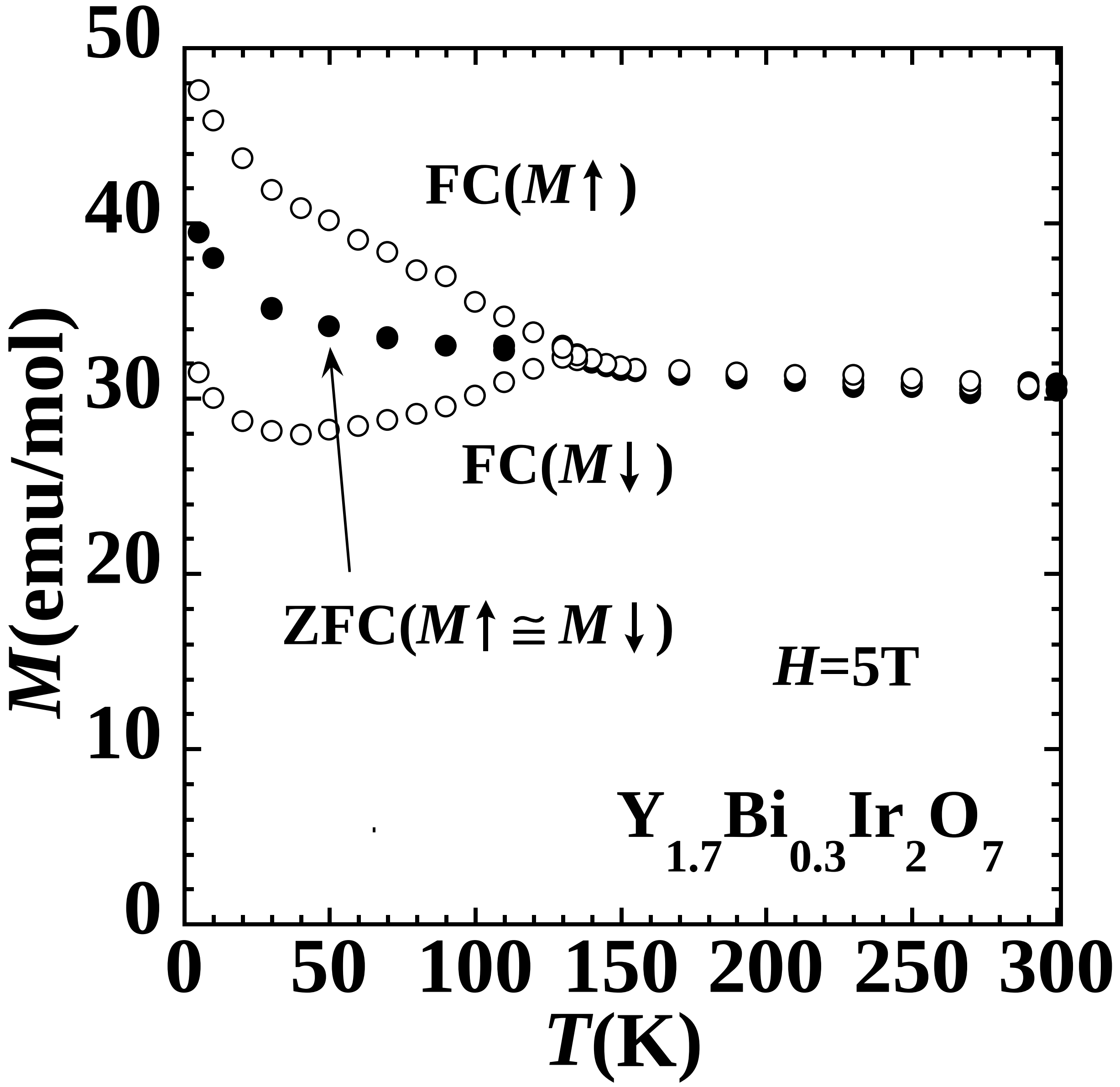

Fig.6

# $Y_{2-x}Bi_xIr_2O_7$

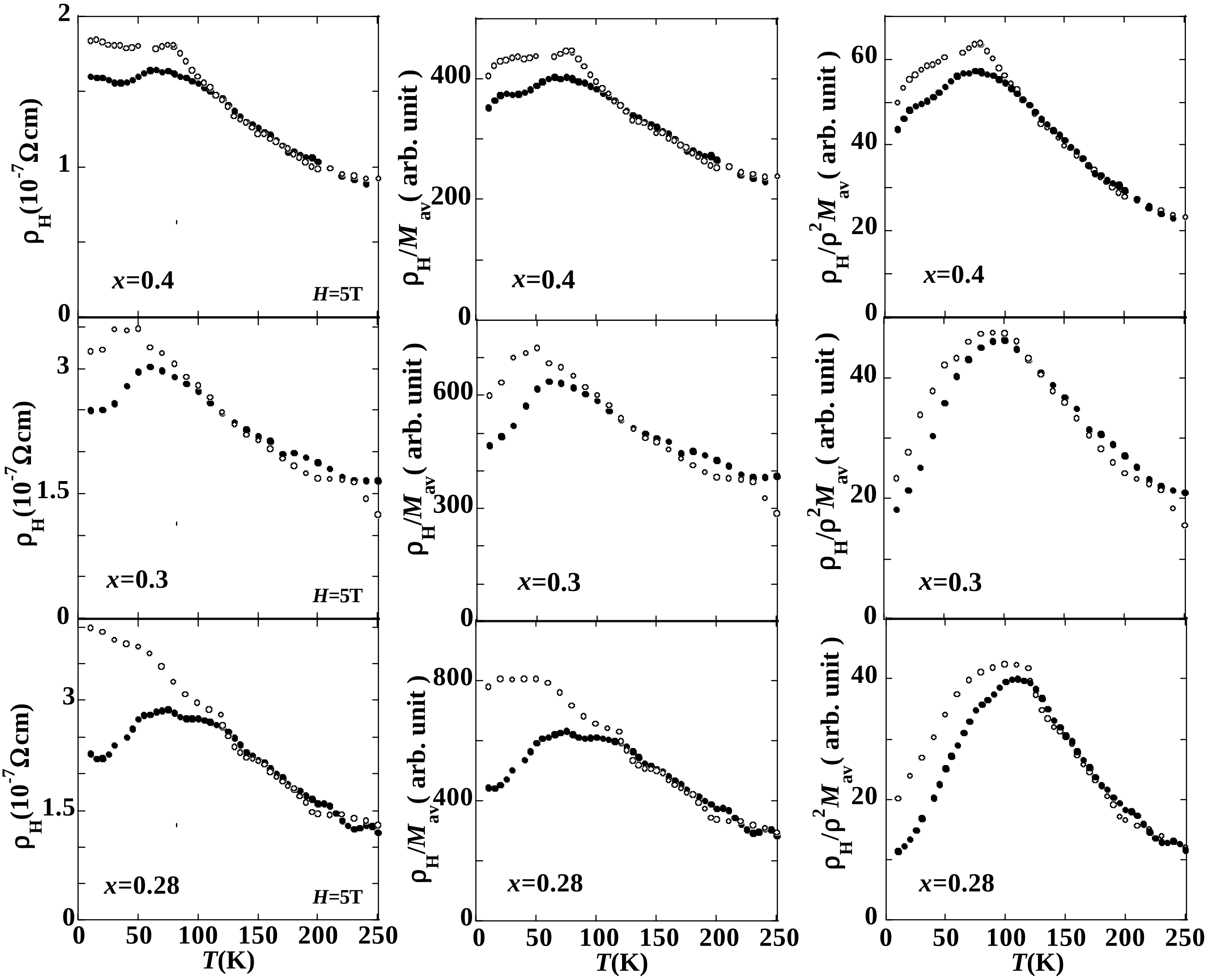

**Fig.7**